\documentclass[english,american,twocolumn,flushbottom,prl]{revtex4-2}
\usepackage[T1]{fontenc}
\usepackage[utf8]{inputenc}
\usepackage{xcolor}
\usepackage{babel}
\usepackage{array}
\usepackage{verbatim}
\usepackage{cprotect}
\usepackage{booktabs}
\usepackage{multirow}
\usepackage{varwidth}
\usepackage{amsmath}
\usepackage{amssymb}
\usepackage{graphicx}
\usepackage[pdfusetitle, bookmarks=true,bookmarksnumbered=false,bookmarksopen=false, breaklinks=false,pdfborder={0 0 1},backref=false,colorlinks=true]{hyperref}
\hypersetup{
linkcolor=magenta,urlcolor=blue,citecolor=blue,pdfstartview={FitH},hyperfootnotes=false,bookmarks=False}
\makeatletter

\date{\today}
\allowdisplaybreaks

\renewcommand{\big}{\bBigg@\@ne}
\renewcommand{\Big}{\bBigg@{1.5}}
\renewcommand{\bigg}{\bBigg@\tw@}
\renewcommand{\Bigg}{\bBigg@{2.5}}

\newcommand{\biggg}{\bBigg@\thr@@}
\newcommand{\Biggg}{\bBigg@{3.5}}

\makeatother
\begin{document}
\title{Incommensuration in odd-parity antiferromagnets}

\author{Changhee Lee}
\email{changhee.lee@otago.ac.nz}
\affiliation{Department of Physics and the MacDiarmid Institute for Advanced Materials
and Nanotechnology, University of Otago, P.O. Box 56, Dunedin 9054,
New Zealand}
\author{Nico A. Hackner}
\affiliation{Department of Physics and the MacDiarmid Institute for Advanced Materials
and Nanotechnology, University of Otago, P.O. Box 56, Dunedin 9054,
New Zealand}
\author{P. M. R. Brydon}
\email{philip.brydon@otago.ac.nz}
\affiliation{Department of Physics and the MacDiarmid Institute for Advanced Materials
and Nanotechnology, University of Otago, P.O. Box 56, Dunedin 9054,
New Zealand}

\begin{abstract}
Inversion-asymmetric antiferromagnets (AFMs) with odd-parity spin-polarization pattern have been proposed as a new venue for spintronics. These AFMs require commensurate ordering to ensure an effective time-reversal symmetry, which guarantees a strictly antisymmetric spin polarization of the electronic states. Recently, non-symmorphic centrosymmetric crystals have been identified as a broad class of materials which could exhibit unit-cell doubling magnetism with odd-parity spin-polarization. 
Here we investigate the stability of these states  against incommensuration. We first demonstrate that the symmetry conditions which permit a $p$-wave spin polarization pattern also permit the existence of a non-relativistic Lifshitz invariant in the phenomenological Ginzburg-Landau free energy. This implies magnetism with an incommensurate ordering vector, independent of its microscopic origin. AFMs with $f$- or $h$-wave spin-polarization are also prone to incommensurability, especially when they have an itinerant origin. Here the symmetry which ensures the odd-parity spin-polarization also guarantees the existence of van Hove saddle points off the time-reversal-invariant momenta, which promote incommensurate spin fluctuations in quasi-two-dimensional electronic systems. Finally, we study the effect of weak spin-orbit coupling in locally noncentrosymmetric materials and find that it favors antiferromagnetic phases with in-plane magnetic moments. However, the inclusion of the spin-orbit coupling also introduces a new mechanism for driving incommensuration. Our results imply that odd-parity AFMs are likely to be preceded by an incommensurate phase, or emerge directly from the normal state via a first order transition. These conclusions are consistent with the phase diagram of several candidate materials.
\end{abstract}
\maketitle

\section{Introduction}

Unconventional magnetism is a rapidly emerging field of study. Much attention has been paid to altermagnets, whose defining characteristic is an exotic spin-polarization pattern at the Fermi surface exhibiting $d$-, $g$-, or $i$-wave
symmetry~\citep{Hayami2019,Yuan2020,Mazin2021,Smejkal2022_Beyond,Smejkal2022_AnomalousHall}. The non-trivial symmetry of the spin-polarization pattern arises from a centrosymmetric collinear antiferromagnetic order which breaks time-reversal symmetry, without the necessity of spin-orbit coupling (SOC). Consequently, altermagnets can exhibit phenomena such as the anomalous Hall effect, which converts the spin currents and the charge currents, as well as be used for giant and tunnelling magnetoresistance device~\citep{GonzalezHernandez2021,Karube2022,Smejkal2022_Magnetoresistance,Smejkal2022_AnomalousHall}. There is strong experimental evidence for altermagnetism in MnTe and KV$_{2}$Se$_{2}$O, with a growing list of candidate materials~\citep{Suyoung2024ALTM,Krempasky2024ALTM,Jiang2025ALTM}. 

Meanwhile, non-collinear and coplanar antiferromagnetic phases are
proposed to exhibit a spin-polarization pattern on the Fermi surface
with $p$-wave symmetry, which are referred to as $p$-wave
antiferromagnets~\citep{Hellenes2024}. In contrast to altermagnets, these magnetic
states, which can be viewed as commensurate helimagnets, are characterized by broken inversion symmetry in the magnetic state, while commensurability allows the restoration of time-reversal symmetry in combination with lattice translations~\citep{Hellenes2024,Brekke2024}. The lack of inversion symmetry enables the electrical control over the non-equilibrium spin population and spin-orbit torque, offering convenient avenues for spintronics device~\citep{Manchon2018RMP,DalDin2024}.

Subsequent studies have extended this class to include coplanar antiferromagnetic
states with $f$- and $h$-wave spin-polarization patterns on the Fermi surface. We collectively refer to these states as odd-parity antiferromagnets,
with each type named after the symmetry of its spin-polarization pattern~\citep{Yue2025,CLee2025}. The Landau phenomenological analysis indicates
that the unit-cell doubling odd-parity antiferromagnetic
phases readily emerge in non-symmorphic systems. This establishes non-symmorphic systems as a promising platform for exploring the unconventional magnetism with odd-parity spin-polarization patterns. In particular, in spite of the scarcity of candidate odd-parity AFM systems compared to the cases of altermagnets, around ten candidate materials such as CeNiAsO are indeed found to have a non-symmorphic space group~\citep{Hellenes2024,Yue2025}.

The commensurability of odd-parity AFMs is crucial to the existence of a true odd-parity spin-polarization pattern. Although a $p$-wave spin-polarization pattern can persist for deviations from a commensurate ordering vector, the spin-polarization and the spin-splitting of the bands are diminished~\citep{Hodt2025}; $f$- and $h$-wave spin-polarization pattern, however, require a commensurate ordering vector. Theoretical studies of the effect of spin-momentum locking due to the the odd-parity spin-polarization on transport~\citep{Brekke2024,Chakraborty2024,Tim2025} and superconductivity
\citep{Fukaya2025Review,Sun2025,Sukhachov2025} have all been carried out under the assumption of a commensurate magnetic ordering. 
From a phenomenological point of view, odd-parity magnets are essentially a form of helical magnetism. Typically, however, helimagnets have an incommensurate ordering vector. In view of the importance of commensurate order for odd-parity AFMs, the stability of unit-cell doubling magnetic order in these system against incommensuration is a critical open question. 

In this work, we show that the inclusion of gradient terms in the phenomenological free energy tends to destabilize the  commensurate phases found in the Landau approach. We first show that the symmetry conditions which ensure a $p$-wave spin-polarization pattern also permit the existence of a Lifshitz invariant~\citep{Landau_Statistical} in the Ginzburg-Landau theory in the \textit{non-relativistic} limit. The presence of the Lifshitz invariant means that the unit-cell doubling vector $\vec{Q}$ is not even a local extrema in the momentum dependent static susceptibility, and so a second-order transition into the unit-cell doubling phase cannot occur. Although incommensurate magnetism is a hallmark of itinerant systems, our results are independent of the microscopic description of the magnetism; in particular, we illustrate our conclusion using a classical Heisenberg model. Secondly, we study the cases of the $f$- and $h$-wave antiferromagnetism in tetragonal systems. Although no symmetry-allowed Lifshitz invariants exist, we find a second-order gradient term with a $d$-wave symmetry. We show that this term can be large enough to induce an incommensuration in a quasi-two-dimensional electron system due to the symmetry-enforced presence of type-II van Hove saddle points. We finally investigate the effect of inversion-symmetric SOC, which is generically present in non-symmorphic systems. Apart from breaking the spin-rotation symmetry, we find that the SOC also introduces a pseudo-Lifshitz invariant which drives incommensurate order. We illustrate this result using a model with Rashba SOC arising from the local breaking of inversion symmetry. We conclude by discussing the implication of our work for the phase diagrams of candidate odd-parity antiferromagnets.

\section{$p$-wave AFM: Symmetry-enforced incommensuration}\label{sec:pwave_lifshitz}

\subsection{Equivalence of the symmetry conditions for a Lifshitz invariant
and a $p$-wave spin-polarization} 

\begin{figure*}
\includegraphics[width=0.95\textwidth]{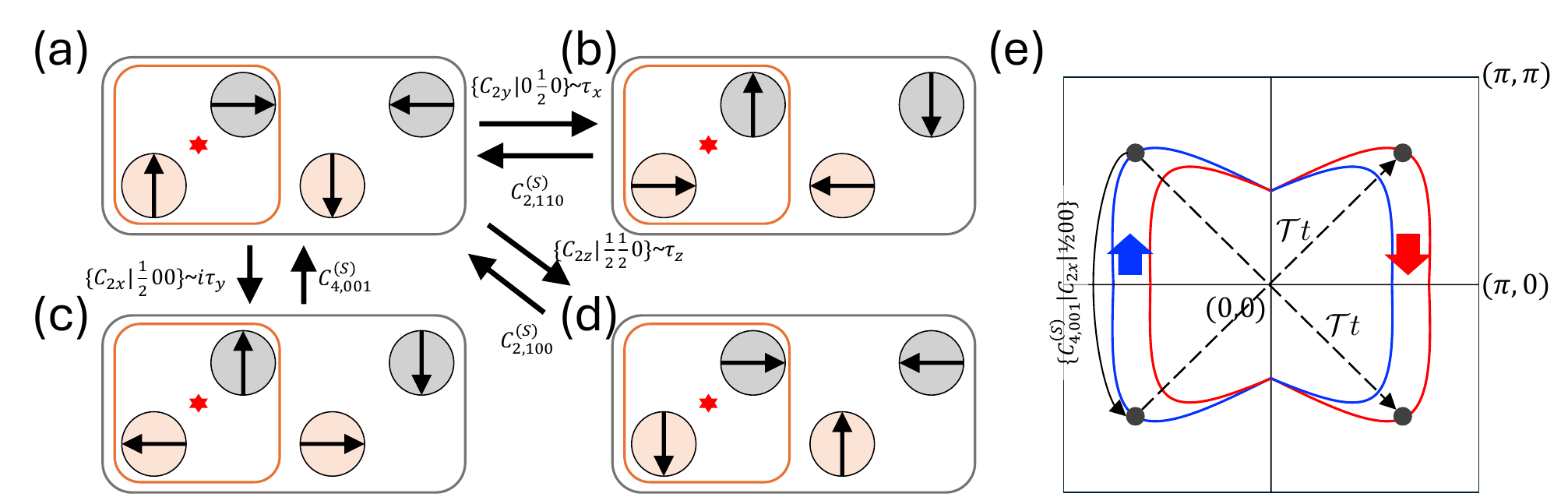}\caption{Spin space group symmetries preserving a non-collinear magnetic state
and the schematics of the corresponding spin-polarization pattern. An antiferromagnetic
state with $\vec{Q}=(\pi,0,0)$, viewed along the $c$-axis, with
magnetic ions occupying the Wyckoff position 2a in the space group
59 is adopted for this illustration. (a-d) Transformation of the non-collinear
coplanar state. The original arrangement of magnetic moments in (a) is mapped into the arrangements shown in (b-d) by the space group symmetries
denoted in the Shubnikov notation $\{C|\vec{\tau}\}$ next to arrow,
and brought back to itself by $m$-fold spin-rotations $C_{m,\hat{n}}^{(S)}$
along the direction $\hat{n}$. The gray and orange boxes depict the
magnetic and the structural unit cells, respectively, while the orange and gray
disks correspond to two sublattices in a structural unit cell. The red star marks
the origin of the coordinate system, which also serves as the inversion
center. (e) Schematics of the corresponding spin-polarization pattern of the electronic Fermi surface in the $k_{z}=0$ plane of the first Brillouin zone of the normal phase. The dashed-line arrows represent the action of the effective time-reversal symmetry ${\cal T}t$. The spin-polarization over the Fermi surface is color-encoded; blue/red represent spin up/down polarization as denoted by arrows. In this illustrative example, each two-fold rotation symmetries of the real space are represented
by $\{C_{2z}|\frac{1}{2}\frac{1}{2}0\}\propto\tau_{z}$, $\{C_{2x}|\frac{1}{2}00\}\propto i\tau_{y}$,
and $\{C_{2y}|0\frac{1}{2}0\}\propto\tau_{x}$.}\label{fig:fig1}
\end{figure*}

We begin by proving the equivalence of the existence of a Lifshitz invariant and the potential appearance of an antiferromagnetic phase with $p$-wave spin-polarization. When the SO(3) spin rotation symmetry is present, the magnetic order parameter can be treated as a scalar order parameter because the spin rotation is decoupled from the lattice rotation~\citep{McClarty2024,CLee2025}. Thus, we present our argument using scalar order parameters but the explicit extension to vectorial order parameters is straightforward.

The Lifshitz invariant is a gradient term in the
Ginzburg-Landau (GL) free energy density which is a bilinear of the order parameters $\phi_{a}$ and their derivatives $\partial_{\nu}\phi_{b}$, specifically
\begin{equation}
{\cal F}_{\text{Lifshitz}}=K_\nu\left(\phi_{a}\partial_{\nu}\phi_{b}-\phi_{b}\partial_{\nu}\phi_{a}\right)\label{eq:Lifshitz}
\end{equation}
where the summation over $\nu=x,y,z$ is understood. Since the exchange $\phi_{a}\leftrightarrow\phi_{b}$
results in the change of the overall sign up to a total derivative, the Lifshitz invariant is allowed to appear in a GL free energy when the antisymmetrized direct
product $\phi_{a}\otimes\phi_{b}(=-\phi_{b}\otimes\phi_{a})$ contains
a vectorial representation to compensate the transformation of $\partial_{\nu}$
under symmetry operations.

Lifshitz invariants appear generally in the GL theory for noncentrosymmetric
systems, where they are typically understood to be derived from relativistic
effects such as the Dzyaloshinskii-Moriya interaction, which originates
from SOC \citep{Togawa2023DMI}. In centrosymmetric
systems, however, the inversion symmetry usually forbids Lifshitz invariants because the irreducible representations (irreps) at time-reversal invariant momentum are typically characterized by a definite parity, i.e., the basis $\phi_{i=1,\cdots,n}$ of a $n$-component irrep have the same parity. In non-symmorphic centrosymmetric space groups, however, there are irreps without a definite parity at certain time-reversal invariant momenta. These mixed-parity irreps allow for a Lifshitz invariant.

When the mixed-parity irreps are constructed by a pair of real-valued
order parameters $\phi_{1}$ and $\phi_{2}$, the symmetry operators
are represented using the real Pauli matrices $\tau_{0},\tau_{x},i\tau_{y},\tau_{z}$.
As the irrep is not characterized by a definite parity, the inversion symmetry
is represented by $\tau_{x}$ or $\tau_{z}$; in the following we take $\tau_{x}$
to represent the inversion without loss of generality. The antisymmetrized
direct product $\phi_{1}\otimes\phi_{2}$ is then odd under the inversion in striking contrast to the cases of symmorphic space groups. Therefore,
the odd parity nature of the antisymmetrized product $\phi_{1}\otimes\phi_{2}$
permits Lifshitz invariants at unit-cell doubling momentum $\vec{Q}$
in non-symmorhpic systems.

Provided the odd-parity antisymmetrized direct product $\phi_{1}\otimes\phi_{2}$, the permitted nonzero values of $K_\nu$ in Eq.~\eqref{eq:Lifshitz} are determined by the representation of other spatial symmetries. Thus, the relevant Lifshitz invariants are associated with the representation matrices in mixed-parity irreps.
We illustrate this association by using a unit-cell doubling ordering
vector $\vec{Q}$ whose little co-group is isomorphic to $D_{{\rm 2h}}$,
which is generated by the inversion and two two-fold rotation symmetries
$C_{2z}$ and $C_{2x}$.

Table~\ref{tab:D2h_tau_x} shows four representative real-valued
mixed-parity irreps at a vector $\vec{Q}$ invariant under $D_{{\rm 2h}}$.
Other mixed-parity irreps are either equivalent to the cases in Table~\ref{tab:D2h_tau_x}
up to the overall sign of representation matrices, a unitary transformation,
or a permutation between the representation matrices for the three
two-fold rotations. The first case in the table corresponds to $C_{2z}\propto\tau_{x}$
and $C_{2x}\propto i\tau_{y}$, and thus $\phi_{1}\otimes\phi_{2}$
is odd and even under $C_{2z}$ and $C_{2x}$, respectively. Since
the symmetry of $\phi_{1}\otimes\phi_{2}$ under two-fold rotation
symmetries is the same with $\partial_{x}$, a Lifshitz invariant
$\phi_{1}\partial_{x}\phi_{2}-\phi_{2}\partial_{x}\phi_{1}$ is allowed
by symmetry. The other cases are obtained by noting that $\phi_{1}\otimes\phi_{2}$
is even (odd) under symmetries represented by $\tau_{0}$ or $i\tau_{y}$
($\tau_{x}$ or $\tau_{z}$). This argument is naturally extended
to the vectorial order parameters $\vec{S}_{1}(\vec{x})$ and $\vec{S}_{2}(\vec{x})$,
representing magnetic orders characterized by the ordering vector
$\vec{Q}$, and $\vec{S}_{1}\cdot\partial_{x}\vec{S}_{2}-\vec{S}_{2}\cdot\partial_{x}\vec{S}_{1}$
is the associated Lifshitz invariant.

\begin{table}
\centering
\begin{tabular}{cccccc}
\toprule 
${\cal I}$ & $C_{2z}$ & $C_{2x}$ & $C_{2y}$ & LI & SPP\tabularnewline
\midrule
\multirow{4}{*}{$\tau_{x}$} & $\tau_{0}(+)$ & $\tau_{z}(-)$ & $\tau_{z}(-)$ & $\partial_{z}$ & $p_{z}$\tabularnewline
 & $\tau_{x}(-)$ & $i\tau_{y}(+)$ & $\tau_{z}(-)$ & $\partial_{x}$ & $p_{x}$\tabularnewline
 & $\tau_{x}(-)$ & $\tau_{z}(-)$ & $i\tau_{y}(+)$ & $\partial_{y}$ & $p_{y}$\tabularnewline
 & $\tau_{0}(+)$ & $i\tau_{y}(+)$ & $i\tau_{y}(+)$ & ($\partial_{x}\partial_{y}\partial_{z}$) & $p_{x}p_{y}p_{z}$\tabularnewline
\bottomrule
\end{tabular}\caption{Lifshitz invariant allowed by mixed-parity physically irreducible
representations at a unit-cell doubling $\vec{Q}$ whose little co-group
is isomorphic to $D_{{\rm 2h}}$ and the spin-polarization pattern
in the associated non-collinear and coplanar antiferromagnetic phase.
The representation matrix for the inversion is fixed as $\tau_{x}$.
The $\pm$ signs in parenthesis represent the parity of the antisymmetrized product $\phi_{1}\otimes\phi_{2}$ under the operation of the symmetry in the corresponding column. In the last row, ($\partial_{x}\partial_{y}\partial_{z}$)
does not mean a Lifshitz invariant but represents that $\phi_{1}\otimes\phi_{2}$
transforms like $xyz$. LI: Lifshitz invariant, SPP: spin-polarization
pattern.}\label{tab:D2h_tau_x}
\end{table}

Given that the mixed-parity irreps responsible for odd-parity AFMs
also permit Lifshitz invariants in the GL free energy, we turn to
establishing the equivalence between the existence of a Lifshitz
invariant and the symmetry conditions necessary for an AFM with $p$-wave spin-polarization pattern. To illustrate their equivalence, let us imagine a lattice constructed by two sublattices of magnetic ions of a single kind, such as shown in Figure~\ref{fig:fig1}(a), which depicts a coplanar AFM state with $\vec{Q}=(\pi,0,0)$ with magnetic ions occupying the Wyckoff position 2a of the orthorhombic
space group 59. The orange and gray disks depict the two sublattices, and the arrows inside the disks represent the magnetic moments.

The spin-polarization pattern of the electronic states in the magnetic
state shown in Fig.~\ref{fig:fig1}(a) is determined by the symmetries
of the magnetic state, some of which are shown in Fig.~\ref{fig:fig1}:
the magnetic state in Fig.~\ref{fig:fig1}(a) is transformed to the
other three states in Figs.~\ref{fig:fig1}(b-d) through two-fold
real-space rotation symmetries denoted next to the black arrows, and
then brought back to its original configuration in Fig.~\ref{fig:fig1}(a)
by the $m$-fold spin-rotation symmetries $C_{m,\hat{n}}^{(S)}$ along
the direction $\hat{n}$. Therefore, the magnetic state shown in Fig.~\ref{fig:fig1}(a) is invariant under, for example, the successive application of the real-space symmetry $\{C_{2x}|\frac{1}{2}00\}$ and the spin-rotation symmetry
$C_{4,001}^{(S)}$. Note that $C_{4,001}^{(S)}$ does not flip the
spin-$z$ component of an electronic state as depicted in Fig.~\ref{fig:fig1}(e).
Taking the effective time-reversal symmetry ${\cal T}t$ as the
usual time-reversal symmetry ${\cal T}$ combined with the unit translation $t=\{e|100\}$, this implies a $p_x$-like spin-polarization pattern, which is odd under the reflection against the $yz$ plane as illustrated in Fig.~\ref{fig:fig1}(e). 

It is worth noting that $\{C_{2x}|\frac{1}{2}00\}$, $\{C_{2z}|\frac{1}{2}\frac{1}{2}0\}$,
and $\{C_{2y}|0\frac{1}{2}0\}$ are represented by $i\tau_{y}$, $\tau_{z}$,
and $\tau_{x}$, respectively, in the example shown in Fig.~\ref{fig:fig1}. Here
$i\tau_{y}$ and $\tau_{z}$ appear due to the alternating behavior
of the unit-cell doubling magnetic order. In addition, it is only
$\{C_{2x}|\frac{1}{2}00\}$ which is combined with a polarization
preserving spin rotation $C_{4,001}^{(S)}$; the other two rotations
are associated with spin rotations that flip the spin-$z$ component
of an electronic state, which is directly related with the $p$-wave
spin-polarization pattern in Fig.~\ref{fig:fig1}(e).

This observation illustrates a general principle about the
relation of the mixed parity irreps and the symmetries in the spin-polarization
pattern in odd-parity AFMs: the space group symmetries represented
by $\tau_{0}$ or $i\tau_{y}$ can be combined with polarizaton preserving
spin-rotations to construct symmetries of the odd-parity AFM state
and determine the spin-polarization pattern of the electronic
states, as demonstrated in Fig.~\ref{fig:fig1}, whereas symmetries represented
by $\tau_{x}$ or $\tau_{z}$ are associated with spin-flipping spin
rotations. The `SPP' column of Table~\ref{tab:D2h_tau_x} summarizes
the possible spin-polarization pattern for each mixed-parity irrep
at $\vec{Q}$ invariant under the $D_{{\rm 2h}}$ point group. All
cases which allow for an antiferromagnetic state with
$p$-wave spin-polarization also allow for a Lifshitz invariant. The last
case with $f$-wave spin-polarization pattern is even under all two-fold rotations, and so does not support a Lifshitz invariant. This situation only occurs at ordering vectors $\vec{Q}=(\pi,\pi,0)$ (space group 59), $\vec{Q}=(0,\pi,\pi)$ (space group 62), and $\vec{Q}=(\pi,\pi,\pi)$ (space groups 58 and 59)~\citep{Suh2023,Yue2025}.

We extend the analysis to other types of $\vec{Q}$, and Table~\ref{tab:D4h_tau_x}
shows the relation between Lifshitz invariants and spin-polarization
pattern for $\vec{Q}$ invariant under $D_{{\rm 4h}}$ (see Appendix~\ref{App:Reality_MixedIrrep}
for other $\vec{Q}$'s). Of the eight cases shown in Table~\ref{tab:D4h_tau_x},
only three cases appear with a Lifshitz invariant with derivative along the four-fold rotation axis, which is chosen as the $z$-axis
in Table~\ref{tab:D4h_tau_x}. Note that these two cases are characterized
by $p$-wave spin-polarization patterns accordingly.

We also want to note that body-centered lattices subject to a symmorphic space group can host odd-parity AFMs due to mixed-parity irreps at non-time-reversal-invariant momentum. For example, the unit-cell quadrupling vector $P(\pi,\pi,\pi)$ in the Brillouin zone of the body-centered tetragonal lattice subject to the space group 139 (I4/mmm) involves physically irreducible representations mixed in parity. Here the body-centering translation $\vec{\tau}=(1/2,1/2,1/2)$ is represented by $i\tau_y$, while the inversion, $C_{2y}$, and $C_{2,110}$ are represented by $\tau_z$, $\tau_0$, $\tau_z$, respectively. Considering that symmetries represented by $\tau_z$ or $\tau_x$ are associated with spin-flipping spin rotations, a coplanar phase with in symmorphic body-centered lattices, such as Sr$_2$CuO$_2$Cu$_2$S$_2$~\citep{Blandy2018}, $\vec{Q}=(\pi,\pi,\pi)$ therefore is expected to exhibit an $f$-wave spin-polarization pattern. 

\begin{table}
\centering
\begin{tabular}{cccccccc}
\toprule 
${\cal I}$ & $C_{2y}$ & $C_{2,110}$ & $C_{4z}$ & $C_{2z}$ & $C_{2x}$ & LI & SPP\tabularnewline
\midrule
\multirow{8}{*}{$\tau_{x}$} & $i\tau_{y}$ & $\tau_{0}$ & $i\tau_{y}$ & $\tau_{0}$ & $i\tau_{y}$ & No & $p_{x}p_{y}p_{z}(p_{x}^{2}-p_{y}^{2})$\tabularnewline
 & $i\tau_{y}$ & $\tau_{x}$ & $\tau_{z}$ & $\tau_{0}$ & $i\tau_{y}$ & No & $p_{x}p_{y}p_{z}$\tabularnewline
 & $i\tau_{y}$ & $\tau_{z}$ & $\tau_{x}$ & $\tau_{0}$ & $i\tau_{y}$ & No & $p_{x}p_{y}p_{z}$\tabularnewline
 & $\tau_{0}$ & $\tau_{z}$ & $\tau_{z}$ & $\tau_{0}$ & $\tau_{0}$ & No & $p_{x}p_{y}p_{z}$\tabularnewline
 & $\tau_{z}$ & $\tau_{0}$ & $\tau_{z}$ & $\tau_{0}$ & $\tau_{z}$ & No & $p_{z}(p_{x}^{2}-p_{y}^{2})$\tabularnewline
 & $\tau_{z}$ & $\tau_{x}$ & $i\tau_{y}$ & $\tau_{0}$ & $\tau_{z}$ & $\partial_{z}$ & $p_{z}$\tabularnewline
 & $\tau_{z}$ & $\tau_{z}$ & $\tau_{0}$ & $\tau_{0}$ & $\tau_{z}$ & $\partial_{z}$ & $p_{z}$\tabularnewline
 & $\tau_{x}$ & $\tau_{z}$ & $i\tau_{y}$ & $\tau_{0}$ & $\tau_{x}$ & $\partial_{z}$ & $p_{z}$\tabularnewline
\bottomrule
\end{tabular}
\cprotect\caption{Lifshitz invariant allowed by mixed-parity irreps at a unit-cell
doubling $\vec{Q}$ whose little co-group is isomorphic to $D_{{\rm 4h}}$
and the spin-polarization pattern in the associated non-collinear
and coplanar antiferromagnetic phase. The representation matrix for
the inversion is chosen as $\tau_{x}$. The overall minus sign, if
it were, are omitted for brevity. ``No'' means there is no Lifshitz invariant, and the symmetry of the antisymmetrized product of order parameter has the same symmetry of the spin-polarization pattern in the last column. LI: Lifshitz invariant, SPP: spin-polarization pattern.}\label{tab:D4h_tau_x}
\end{table}

\subsection{Incommensurate phase by\\Lifshitz invariants of non-relativistic origin}

Given the existence of Lifshitz invariant, the GL free energy describing
the transition from a normal phase to an antiferromagnetic state with
a $p$-wave spin-polarization pattern is
written as
\begin{align}
{\cal F}_{{\rm GL}}= & \int_{\vec{r}}{\cal F}_{{\rm L}}+K_{\nu}(\vec{S}_{2}\cdot\partial_{\nu}\vec{S}_{1}-\vec{S}_{1}\cdot\partial_{\nu}\vec{S}_{2})+{\cal O}(\partial_{\nu}^{2}),\label{eq:GLfree}\\
{\cal F}_{{\rm L}}= & \alpha(T)(\vec{S}_{1}^{2}+\vec{S}_{2}^{2})+\beta_{1}(\vec{S}_{1}^{2}+\vec{S}_{2}^{2})^{2}+\beta_{2}(\vec{S}_{1}\cdot\vec{S}_{2})^{2} \nonumber\\
&
+\beta_{3}|\vec{S}_{1}|^{2}|\vec{S}_{2}|^{2},
\end{align}
where  the space-dependence
of $\vec{S}_{i}(\vec{r})$ is omitted for conciseness. Here ${\cal F}_L$ is the Landau free energy previously obtained in~\cite{Yue2025,CLee2025}. The local magnetic moment is expressed as $\vec{m}(\vec{r})=\sum_{i}{\rm Re}[\vec{S}_{i}(\vec{r})e^{i\vec{Q}\cdot\vec{r}}]$, which describes a simple unit-cell doubling order when $\vec{S}_{i}(\vec{r})$ is constant. Taking the Fourier transform $\vec{S}_{i,\vec{q}}=\int_{\vec{q}}\vec{S}_{i}(\vec{r})e^{-i\vec{q}\cdot\vec{r}}$,
the GL free energy in the momentum space is written as
\begin{equation}
{\cal F}_{{\rm GL}}\approx\int_{\vec{q}}(\vec{S}_{1,\vec{q}}^{*},\vec{S}_{2,\vec{q}}^{*})\bigg(\begin{matrix}\alpha(T) & -iK_{\nu}q_{\nu}\\
iK_{\nu}q_{\nu} & \alpha(T)
\end{matrix}\bigg)\bigg(\begin{matrix}\vec{S}_{1,\vec{q}}\\
\vec{S}_{2,\vec{q}}
\end{matrix}\bigg),\label{eq:MomentumGL_w_Lifshitz}
\end{equation}
where we keep only terms quadratic in the order parameters and to linear order in $\vec{q}$. The Lifshitz terms appear in the off-diagonal antisymmetric part of the matrix in Eq.~\eqref{eq:MomentumGL_w_Lifshitz}. A continuous phase transition occurs when the lowest eigenvalue $\alpha_{-}(\vec{q})=\alpha(T)-|K_{\nu}q_{\nu}|+{\cal O}(q^{2})$
of the matrix in Eq.~\eqref{eq:MomentumGL_w_Lifshitz} becomes zero. Due to the term $|K_{\nu}q_{\nu}|$ linear in $\vec{q}$, we always have $\alpha_{-}(\vec{q})<\alpha(T)$, which means that the minimum of $\alpha_{-}(\vec{q})$ is always found at $\vec{q}\neq0$, and thus the resultant phase is expected to be incommensurate. Consequently, the symmetry which allows the $p$-wave spin polarization pattern of the unit-cell-doubling AFM also implies that an incommensurate phase has a higher transition temperature.

We emphasize that the Lifshitz invariant in our theory is non-relativistic in origin as it is permitted in the non-relativistic spin group, and thus may be expected to be large even in the absence of heavy elements.

\begin{figure}
\includegraphics[width=0.65\columnwidth]{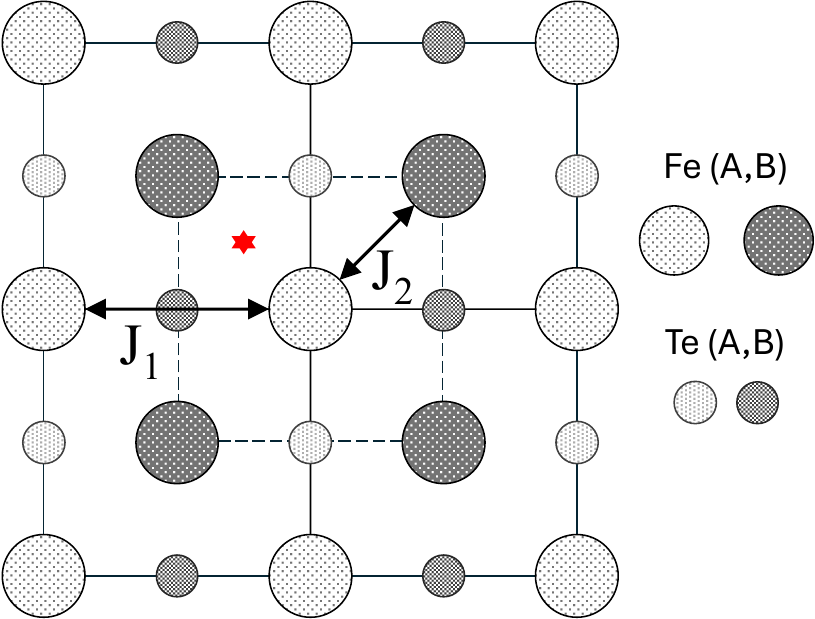}\caption{Intrasublattice and intersublattice exchange interactions between
iron atoms in the tetragonal FeTe (space group 129, P4/nmm). For conciseness, the nearest intra- and inter-sublattice interactions $J_{1}$ and $J_{2}$, respectively, in a plane of iron atoms are depicted.}\label{fig:FeTe}
\end{figure}

Since it is based purely on symmetry, our argument is not limited to the GL phenomenological description of an itinerant magnetic transition, but also applies to a local moment scenario. Here we illustrate this using a classical Heisenberg model. Specifically, we consider a lattice which consists of two magnetic ions in a structural unit cell such as shown in Figure~\ref{fig:FeTe} for a plane in FeTe. In the classical limit, the Heisenberg model Hamiltonian is written in the momentum space as 
\begin{align}
\hat{H}= & \sum_{\vec{k}}\big(\vec{s}_{A}(-\vec{k}),\vec{s}_{B}(-\vec{k})\big)J(\vec{k})\bigg(\begin{matrix}\vec{s}_{A}(\vec{k})\\
\vec{s}_{B}(\vec{k})
\end{matrix}\bigg),\label{eq:spinHam_Fourier}
\end{align}
with
\begin{equation}
J(\vec{k})=\left(\begin{matrix}J_{AA}(\vec{k}) & J_{AB}(\vec{k})\\
J_{AB}(-\vec{k}) & J_{BB}(\vec{k})
\end{matrix}\right),
\end{equation}
in terms of the spin operators $\vec{s}_{i}(\vec{k})=\sum_{\vec{R}}s_{i}(\vec{R})e^{-i\vec{k}\cdot(\vec{R}+\vec{r}_{i})}$.
$\vec{R}$ labels the lattice vectors and $\vec{r}_{i=1,2}$ denote
the position of two sublattices occupied by magnetic ions in a structural
unit cell. $J_{AA(BB)}(\vec{k})$ and $J_{AB}(\vec{k})$ represent
the Fourier transform of the intrasublattice and the intersublattice exchange interactions, respectively. According to the Luttinger-Tisza-Lyon-Kaplan method~\citep{Lyons1960LTLKMethod,Schmidt2022LTLKMethod,Rajeev2025LTLKMethod}, the wavevector of the stable spiral state corresponds to the wavevector which gives the most negative eigenvalue of $J(\vec{k})$.

The effect of symmetry on the stability of the continuous phase transition into a
unit-cell doubling $p$-wave AFM compared to an incommensurate ordering is examined by expanding the matrix $J(\vec{k})$ in this model around a unit-cell doubling vector
$\vec{k}=\vec{Q}$, where a mixed-parity irrep should exist at $\vec{Q}$ to permit a continuous phase transition into a
$p$-wave AFM. The symmetry properties of $\vec{s}_{i}(\vec{Q})$
dictated by the mixed-parity irrep indicates that the expansion of
$J(\vec{Q}+\vec{q})$ with respect to $\vec{q}$ takes the following
form:
\begin{equation}
J(\vec{Q}+\vec{q})\approx\left(\begin{matrix}J_{AA}(\vec{Q}) & -i\vec{q}\cdot\vec{K}_{AB}(\vec{Q})\\
i\vec{q}\cdot\vec{K}_{AB}(-\vec{Q}) & J_{AA}(\vec{Q})
\end{matrix}\right),\label{eq:J(Q+q)}
\end{equation}
with $\vec{K}{}_{AB}(\vec{Q})=i\nabla_{\vec{k}}J_{AB}(\vec{k})\big|_{\vec{k}=\vec{Q}}$.
Here, $J_{AA}(\vec{Q})=J_{BB}(\vec{Q})$, $J_{AB}(\vec{Q})=0$, and
the presence of the off-diagonal elements linear in $\vec{q}$ is allowed by the mixed-parity symmetry of $\vec{s}_{i}(\vec{Q})$.
As a result, Eq.~(\ref{eq:J(Q+q)}) is essentially has the same form as the matrix in Eq.~(\ref{eq:MomentumGL_w_Lifshitz}). In particular, the smallest eigenvalue of $J(\vec{Q}+\vec{q})$ is $\lambda_{-}(\vec{Q}+\vec{q})=J_{{\rm AA}}(\vec{Q})-|\vec{q}\cdot\vec{K}_{AB}(\vec{Q})|<\lambda_-(\vec{Q})$, and thus $\lambda_{-}(\vec{k})$ does not have even a local minima at $\vec{Q}$ due to the linear-in-$\vec{q}$ term. Consequently,
a continuous phase transition into a state with $\vec{k}=\vec{Q}$ from the normal phase is disallowed by the non-relativistic
intersublattice exchange interaction $J_{AB}$.

Figure~\ref{fig:FeTe} illustrates a plane of iron atoms in the tetragonal FeTe. $J_1$ and $J_2$ denote the nearest-neighbor intra- and inter-sublattice exchange interactions in the plane, respectively. Other exchange interactions are omitted for conciseness as they do not qualitatively change the following argument, but may be necessary to stabilize an AFM state with wavevector close to a unit-cell doubling value. In our example we have $J_{AA}(\vec{k})=J_{BB}(\vec{k})=J_{1}(\cos k_{x}+\cos k_{y})$
and $J_{AB}(\vec{k})=J_{2}\cos(k_x /2)\cos(k_y /2)$, and thus we obtain
$J_{AB}(\pm\vec{Q}\pm\vec{q})=-J_{2}\cos(q_x /2)\sin(q_y/2)\approx-J_2q_{y}/2$
for $\vec{Q}=(0,\pm\pi,0)$ as well as $\vec{Q}=(0,\pm\pi,\pi)$,
which gives finite linear-in-$\vec{q}$ terms in Eq.~\eqref{eq:J(Q+q)}. This symmetry-allowed linear-in-$\vec{q}$ exchange term between the neighboring Fe atoms implies that fluctuations at an incommensurate momentum $\vec{Q}+\vec{q}$ stronger than that at $\vec{Q}=(0,\pi,\pi)$, which is indeed observed in the inelastic neutron scattering in Fe$_{1+y}$Te~\citep{Parshall2012FeTe}. In this system the unit-cell doubling antiferromagnetism with $\vec{Q}=(0,\pi,\pi)$ eventually emerges through a first order transition.

Concerning the cases with more than two magnetic sublattices in a
structural unit cell, the same conclusion can be drawn by noting that
the matrix $J(\vec{Q})$ is block-diagonalized with 2 by 2 subblocks
each of which takes the form of Eq.~(\ref{eq:J(Q+q)}).
To conclude this subsection, we note that the fragility of the $p$-wave AFM to incommensuration demonstrated here is also a feature of conventional heli- or spiral-magnets, in the sense that ordering vectors at time-reversal invariant momenta do not correspond to minima of the free energy.

\subsection{Spin-polarization pattern in incommensurate phase}

Here we investigate the Landau free energy of incommensurate order parameters which are favoured by the Lifshitz invariant. For simplicity, we assume the
incommensurate ordering vector $\vec{k}=\vec{Q}+\vec{q}$ is located
inside the first Brillouin zone, which is relevant when $\vec{Q}$
is one of the points $(\pi,0,0)$, $(0,\pi,0)$, and $(0,0,\pi)$
of the Brillouin zone of orthorhombic, tetragonal, and cubic systems,
or when $\vec{Q}=(0,0,\pi)$ in trigonal and hexagonal systems.

For the orthorhombic case, the relevant Landau free energy for the
incommensurate order parameters with highest transition temperature
is written as
\begin{align}
F & =\alpha_{\vec{k}}(T-T_{c})(\vec{M}_{1,\vec{k}}^{2}+\vec{M}_{2,\vec{k}}^{2})+\beta_{1}(\vec{M}_{1,\vec{k}}^{2}+\vec{M}_{2,\vec{k}}^{2})^{2}\nonumber \\
 & +\beta_{2}(\vec{M}_{1,\vec{k}}\cdot\vec{M}_{2,\vec{k}})^{2}-\beta_{2}\vec{M}_{1,\vec{k}}^{2}\vec{M}_{2,\vec{k}}^{2},
\end{align}
which describes the transitions into an antiferromagnetic state characterized
by $\vec{S}_{i}(\vec{R})=\vec{M}_{1,\vec{k}}\cos(\vec{k}\cdot(\vec{R}+\vec{r}_{i}))-\vec{M}_{2,\vec{k}}\sin(\vec{k}\cdot(\vec{R}+\vec{r}_{i}))$
where $\vec{R}$ labels the structural unit cells and $\vec{r}_{i=A,B}$
denote the position of two sublattices in the physical unit cell.
This Landau free energy for the incommensurate order is thermodynamically
stable when $\beta_{1}>0$ and $\beta_{2}>-4\beta_{1}$.

For $\beta_{2}<0$, a collinear phase $(\vec{M}_{1,\vec{k}}\parallel\vec{M}_{2,\vec{k}})$
occurs. Due to the incommensurability of the ordering vector, the half-collinear
state and the collinear state discussed in Refs.~\citep{CLee2025,Yue2025}
are not distinguished. A non-collinear coplanar $(\vec{M}_{1,\vec{k}}\perp\vec{M}_{2,\vec{k}})$
solution is found for $\beta_{2}>0$ which gives rise to a spin-polarization
pattern in the electronic structure, which takes the $p$-wave symmetry
since the little co-group of the set $\{\vec{k},-\vec{k}\}$ is $D_{{\rm 2h}}$.

The argument given above is also applicable to the case of $\vec{k}=(0,0,k_{z})$
in the trigonal and hexagonal systems. For tetragonal and cubic systems, however,
the increased multiplicity of the star of incommensurate $\vec{k}$
leads to more complicated free energy; the tetragonal case has
been studied in research on the spin-density wave phase of iron-based
superconductors~\citep{Brydon2011,Christensen2018}. 

For $\vec{Q}$ on the boundaries of the Brillouin zone, the weak form of the Lifshitz condition for the thermodynamic stability of the state with the ordering vector $\vec{Q}$ may not be satisfied and thus the ordering vector of the phase after the transition could be located at a place different from the prediction by the Lifshitz invariant at a high-symmetry point~\citep{Michelson1978,Stokes1993}. However, the presence of a Lifshitz invariant still implies that the phase transition leads to an incommensurate phase.

\section{Incommensuration in $f$- and $h$-wave AFMs} \label{sec:fwave}

Tables~\ref{tab:D2h_tau_x}~and~\ref{tab:D4h_tau_x} also reveal
that AFMs with $f$- and $h$-wave spin-polarization pattern are stable against the incommensuration discussed above. However,
the simultaneous existence of mixed-parity irreps at multiple unit-cell
doubling momenta can still give rise to a strong tendency towards
incommensurate order, especially in itinerant magnets, This can be understood through second-order gradient terms in the GL free
energy
\begin{align} \label{eq:fwaveGLfree}
{\cal F}_{{\rm GL}}= & {\cal F}_{L}-D(\vec{S}_{1}(\vec{r})\cdot\nabla^{2}\vec{S}_{2}(\vec{r})+\vec{S}_{2}(\vec{r})\cdot\nabla^{2}\vec{S}_{2}(\vec{r}))\\
 & -K_{\mu\nu}(\vec{S}_{1}(\vec{r})\cdot\partial_{\mu}\partial_{\nu}\vec{S}_{2}(\vec{r})+\vec{S}_{2}(\vec{r})\cdot\partial_{\mu}\partial_{\nu}\vec{S}_{1}(\vec{r})),\nonumber 
\end{align}
where  ${\cal F}_{L}$ is again the Landau free energy without the gradient terms.
By the symmetry properties of $\vec{S}_{i}(\vec{r})$ enumerated in Tables~\ref{tab:D2h_tau_x}~and~\ref{tab:D4h_tau_x},
the nonzero elements of $K_{\mu\nu}$ are found to be either $K_{xx}=-K_{yy}=K\neq0$ or
$K_{xy}=K_{yx}=K\neq0$. In the momentum space representation, the
lowest eigenvalue of the coefficient matrix of the bilinears of order
parameters is $D(q_{x}^{2}+q_{y}^{2})-2|Kq_{x}q_{y}|$ or $D(q_{x}^{2}+q_{y}^{2})-|K||q_{x}^{2}-q_{y}^{2}|$, respectively.
This informs us of that the $\vec{q}=0$ point becomes a saddle point
when $0<D<|K|$, and the transition would then be expected to occur at a finite
$\vec{q}$. 

In the following, we investigate how symmetry affects the likelihood
of $D<|K|$ in the prototypical scenario of itinerant magnetism driven
by the nesting of Van Hove saddle points (VHSs).

\subsection{Symmetry-assisted incommensuration\\led by Type-II VHSs}\label{sub:VHS}

The VHS-driven magnetism is typically conceived in quasi two-dimensional
systems with tetragonal or hexagonal crystal systems, and we concentrate
on the tetragonal case where the VHSs located around the unit-cell
doubling momentum $(\pi,0,0)$ and $(0,\pi,0)$ can promote an antiferromagnetic
instability with the ordering vector around the unit-cell doubling
$(\pi,\pi,0)$.

As we are interested in the cases where an odd parity AFM with $f$- or $h$-wave spin-polarization pattern can emerge, we suppose the existence of a mixed-parity
irrep at $(\pi,\pi,0)$. This implies the existence of a symmetry
of the normal phase accompanied by a half-translation $\vec{\tau}$
satisfying $(1,1,0)\cdot(2\vec{\tau})\in2\mathbb{Z}+1$~\citep{CLee2025}.
Considering that the half-translation associated with the two-fold
rotation $C_{2,110}$ in tetragonal system takes the form $(a,a,c)$
with $a,c\in\{0,1/2\}$, $\vec{\tau}$ in $\{C_{2y}|\vec{\tau}\}$
should be one of $(1,0,c)$ or $(0,1,c)$ with $c\in\{0,1/2\}$. This
automatically guarantees the existence of mixed-parity irreps at $(\pi,0,0)$
and $(0,\pi,0)$ in tetragonal systems, which correspond to the first
three cases in Table~\ref{tab:D2h_tau_x} with the replacement $y\leftrightarrow z$
understood.

Given the presence of mixed-parity irreps at $(\pi,0,0)$ and $(0,\pi,0)$,
we can use Table~\ref{tab:D2h_tau_x} to construct $\vec{k}\cdot\vec{p}$
Hamiltonians for electrons. In this respect, it is worthy of noting
that the creation/annihilation operators $\hat{c}^{\dagger}$ and
$\hat{c}$ for electrons transform according to the irreducible representations
of single-valued space groups when the SOC is ignored.
This establishes the identification of the symmetry properties of
the bilinears $\hat{c}_{\vec{k},i}^{\dagger}\hat{c}_{\vec{k},j}$
and the bilinears of order parameters $\phi_{\vec{k},i}^{*}\phi_{\vec{k},j}$~\citep{Bradley2009,Onodera1978}. 

\begin{figure*}
\includegraphics[width=0.95\linewidth]{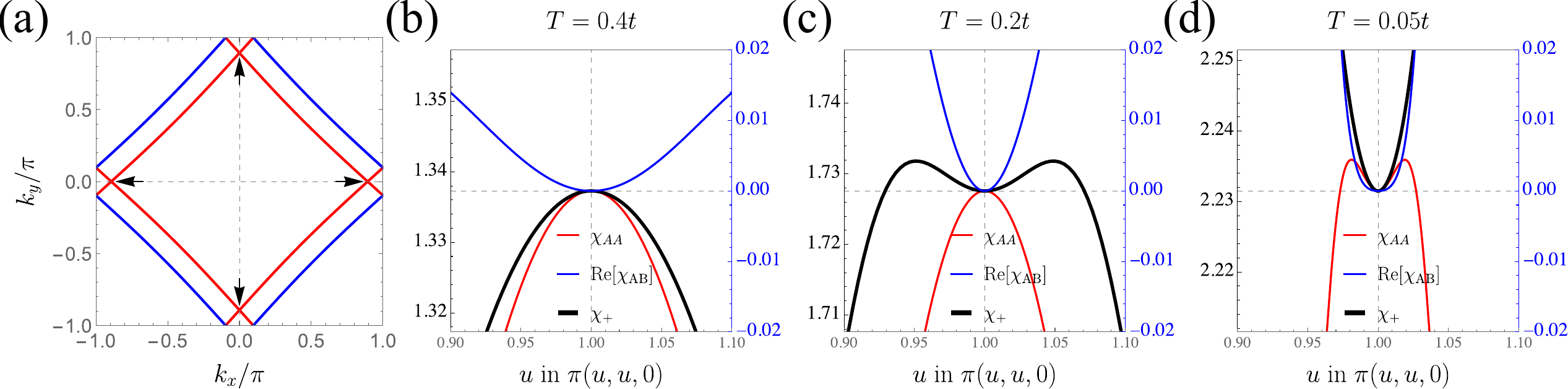}\caption{(a) Fermi surface at $k_{z}=0$ exhibiting the symmetry-enforced
VHSs indicated by black arrows around $\vec{Q}=(\pi,0,0)$ and $\vec{Q}=(0,\pi,0)$.
(b-d) $\chi_{AA}(\vec{k})$ ,${\rm Re}[\chi_{AB}(\vec{k})]$, and
$\chi_{+}(\vec{k})=\chi_{AA}(\vec{k})+|{\rm Re}[\chi_{AB}(\vec{k})]|$
along the high symmetry line $\vec{k}=(u,u,0)$ around $(\pi,\pi,0)$.
For this illustration, $\varepsilon_{\vec{k},0}=-2t(\cos k_{x}+\cos k_{y})+4t'\cos k_{x}\cos k_{y}-\mu$
and $t_{\vec{k},x}-it_{\vec{k},y}=t_{\perp}(1+e^{-ik_{x}})(1+e^{-ik_{y}})$
are used with $(t',t_{\perp},\mu)=(0.05,0.3,-0.1)t$. The van Hove
singularity levels are chosen as the Fermi level $\mu$.}\label{fig:Driven_By_vHS}
\end{figure*}

As noted before, the first three cases in Table~\ref{tab:D2h_tau_x}
are relevant to the momentum $(\pi,0,0)$ and $(0,\pi,0)$, and we
find three kinds of $\vec{k}\cdot\vec{p}$ Hamiltonians: 
\begin{align}
H_{kp} & =\sum_{\vec{q}}\hat{C}_{\vec{q}}^{\dagger}\{\varepsilon_{0}(\vec{q})\tau_{0}+t_{x}f(\vec{q})\tau_{x}+t_{y}q_{y}\tau_{y})\hat{C}_{\vec{q}},\label{eq:Hkp}
\end{align}
where $\varepsilon_{0}(\vec{q})=aq_{x}^{2}+bq_{y}^{2}$ and $f(\vec{q})$
is one of $q_{y}q_{z}$, $q_{x}q_{z}$, and $q_{x}q_{y}$. $\hat{C}_{\vec{q}}=(\hat{c}_{\vec{q},1},\hat{c}_{\vec{q},2})^{T}$
denote the spinless fermionic operators transforming according to
a mixed parity irrep. Since spinless fermions and order parameters
obey the same symmetries, the appearance of the antisymmetric matrix
$\tau_{y}$ with a $q_{y}$-linear coefficient is consistent with
the presence of a Lifshitz invariant for magnetic order at $(0,\pi,0)$. Because of this $\vec{q}$-linear
term, the eigenenergy of $H_{kp}$ in Eq.~(\ref{eq:Hkp}) acquires
a linear dispersion around $\vec{q}=0$:
\begin{equation}
\varepsilon_{\pm}(\vec{q})\approx aq_{x}^{2}+bq_{y}^{2}\pm|t_{y}q_{y}|.
\end{equation}
Accordingly, if $ab<0$, the saddle point normally expected at the TRIM is displaced to nonzero wavevector $\vec{q}\neq0$. Such VHSs off time-reversal momentum are referred to as type-II VHSs~\citep{Yao2015,CLee2025RG}.
Consequently, we have type-II VHSs at $(\pi\pm\delta,0,0)$ and $(0,\pi\pm\delta,0)$
with $\delta\neq0$ whenever a mixed-parity irrep exists at $(\pi,\pi,0)$. As such, the
the nesting vectors $(\pi\pm\delta,\pi\pm\delta,0)$ between the two
VHSs are generally incommensurate.

The consequence of the type-II VHSs connected via a incommensurate nesting vector can be investigated by using the Hubbard model with the tight-binding Hamiltonian for generic two-sublattice systems:
\begin{align}
\hat{H} & =\hat{H}_{{\rm TB}}+U\sum_{\vec{R}}\sum_{\tau=A,B}\hat{n}_{\vec{R},\tau,\uparrow}\hat{n}_{R,\tau,\downarrow},\label{eq:HubbardModel}\\
\hat{H}_{{\rm TB}} & =\sum_{\vec{k}}\hat{C}_{\vec{k}}^{\dagger}\{\varepsilon_{\vec{k},0}\tau_{0}+\varepsilon_{\vec{k},z}\tau_{z}+\vec{t}_{\vec{k}}\cdot\vec{\tau}\}\hat{C}_{\vec{k}},\label{eq:TBHam}
\end{align}
with $\hat{n}_{\vec{R},\tau,\sigma}=\hat{c}_{\vec{R},\tau,\sigma}^{\dagger}\hat{c}_{\vec{R},\tau,\sigma}$ and the trivial (non-trivial) intra-sublattice hopping $\varepsilon_{\vec{k},0}$ ($\varepsilon_{\vec{k},z}$) and the inter-sublattice hoppings $\vec{t}_{\vec{k}}=(t_{\vec{k},x},t_{\vec{k},y})$. $\varepsilon_{\vec{k},z}=0$ when inversion is a site-transposing symmetry. $\hat{c}_{\vec{R},\tau,\sigma}$ is the annihilation operator for an electron with spin $\sigma=\uparrow,\downarrow$ at sublattice $\tau=A,B$ in the unit cell labelled as $\vec{R}$. The Pauli matrices $\tau_{0,x,y,z}$ act on the two sublattices degrees of freedom, which is the minimal number of sublattices in non-symmorphic systems~\citep{CLee2025}. $\hat{C}_{\vec{k}}=(\hat{c}_{\vec{k},A,\uparrow},\hat{c}_{\vec{k},A,\downarrow},\hat{c}_{\vec{k},B,\uparrow},\hat{c}_{\vec{k},B,\downarrow})^{T}$
are the annihilation operators in the momentum space. SOC is ignored in the non-relativistic limit. Figure~\ref{fig:Driven_By_vHS}(a) illustrates representative Fermi surfaces of Eq.~(\ref{eq:HubbardModel})
in the $k_{z}=0$ plane with the Fermi level positioned at the van
Hove singularity. The type-II VHSs off the time-reversal momentums
$(\pi,0,0)$ and $(0,\pi,0)$ appear in accordance with the $\vec{k}\cdot\vec{p}$
theory in Eq.~(\ref{eq:Hkp}).

The magnetic instability in this model can be examined by decomposing
the Hubbard interaction through the spin-rotation-preserving Hubbard-Stratonovich
transformation into spin channels~\cite{CLee2025}, which introduces auxiliary bosonic
fields $\vec{S}_{\tau}(\vec{k})$ representing the Fourier transform
of the magnetization in the sublattice $\tau$. The standard field
theoretic procedure gives the quadratic terms in the Landau free energy
\begin{equation}
\sum_{\tau,\tau'=A,B}\sum_{\vec{k}}[\frac{\delta_{\tau,\tau'}}{U}-\chi_{\tau,\tau'}(\vec{k})]\vec{S}_{\tau}(-\vec{k})\cdot\vec{S}_{\tau'}(\vec{k}).
\end{equation}
The elements of the susceptibility matrix $\chi(\vec{k})$ are given by 
\begin{align}
\chi_{AA}(\vec{k})= & \chi_{BB}(\vec{k})=\frac{1}{2}\sum_{\vec{p}}\sum_{\lambda,\lambda'=\pm}\chi_{\lambda,\lambda'},\\
\chi_{AB}(\vec{k})= & \chi_{BA}^{*}(\vec{k})=\frac{1}{2}\sum_{\vec{p}}\sum_{\lambda,\lambda'=\pm}\frac{t_{\vec{p}}t_{\vec{p}+\vec{k}}^{*}\lambda\lambda'}{|t_{\vec{p}}||t_{\vec{p}+\vec{k}}|}\chi_{\lambda,\lambda'},
\end{align}
where we have the usual Lindhard function
\begin{align}
\chi_{\lambda\lambda'}\equiv & \frac{\tanh\frac{\xi_{\vec{p},\lambda}}{2T}-\tanh\frac{\xi_{\vec{p}+\vec{k},\lambda'}}{2T}}{\xi_{\vec{p},\lambda}-\xi_{\vec{p}+\vec{k},\lambda'}}, \label{eq:lindhard}
\end{align}
while $t_{\vec{p}}=t_{\vec{p},x}-it_{\vec{p},y}$ and $t_{\vec{p}}t_{\vec{p}+\vec{k}}^{*}=\vec{t}_{\vec{p}}\cdot\vec{t}_{\vec{p}+\vec{k}}+i(\vec{t}_{\vec{p}}\times\vec{t}_{\vec{p}+\vec{k}})_{z}$.
At $\vec{k}=(\pi,\pi,0)$, $\chi_{AB}(\vec{k})=0$ so that two order
parameters $\vec{S}_{A}(\vec{k})$ and $\vec{S}_{B}(\vec{k})$ become
degenerate and form a two-dimensional mixed-parity irrep. For the
case without a Lifshitz invariant at $\vec{k}=(\pi,\pi,0)$ in Table
\ref{tab:D4h_tau_x}, the mixed parity irreps dictate $\vec{t}_{\vec{p}}\cdot\vec{t}_{\vec{p}+(\pi,\pi,0)}$
in the integrand of $\chi_{AB}(\vec{k})$ to transform like the $d_{xy}$
or the $d_{x^{2}-y^{2}}$ spherical harmonic, which indicates $\chi_{AB}(\pi+q_{x},\pi+q_{y},0)=2K_{xy}q_{x}q_{y}$
or $K_{xx}(q_{x}^{2}-q_{y}^{2})$, respectively. Meanwhile the cross product $\vec{t}_{\vec{p}}\times\vec{t}_{\vec{p}+(\pi,\pi,0)}$ transforms following the $f$-wave symmetry of the spin-polarization pattern and it does not contribute to second-order gradient terms. We note that this term gives rise to a Lifshitz invariant if we take $k=(0,\pi,0)$. Lastly, we have $\chi_{AA}(\vec{k})=\chi_{BB}(\vec{k})=\chi_{AA}(\pi,\pi,0)-D(q_{x}^{2}+q_{y}^{2})$.

Figs.~\ref{fig:Driven_By_vHS}(b-d) illustrate $\chi_{AA}(\vec{k})$, ${\rm Re}[\chi_{AB}(\vec{k})]$, and the largest eigenvalue $\chi_{+}(\vec{k})=\chi_{AA}(\vec{k})+|{\rm Re}[\chi_{AB}(\vec{k})]|$ along the high-symmetry line $\vec{k}=(u,u,0)$ for three temperatures. Note that $\chi_{+}$ approximates to the largest eigenvalue of the matrix $\chi(\vec{k})$. The right y-axis of each panel represents the value of ${\rm Re}[\chi_{AB}(\vec{k})]$. The negative curvature of $\chi_{AA}(\vec{k})$ around $u=\pi$ corresponds to the usual positive $D$. As shown in Fig.~\ref{fig:Driven_By_vHS}(b), $\chi_{+}(\vec{k})$ is maximized at $(\pi,\pi,0)$ when temperature is larger than the intersublattice hopping amplitude $t_{\perp}$. At lower temperature, $(\pi,\pi,0)$ becomes a saddle point as $|K_{xy}|$ takes over $D$ and drives for $\chi_+$ to have a maximum off the unit-cell doubling point as shown in Fig.~\ref{fig:Driven_By_vHS}(c). We also observe the change of the sign of $D$ in Fig.~\ref{fig:Driven_By_vHS}(d), which reflects the dominance of the nesting between VHSs in low temperature. This makes $|K_{xy}|>D$ making it easier to satisfy the inequality $|K_{xy}|>D$.

We would like to emphasize that the tendency towards an incommensurate
phase can be suppressed if $|\vec{t}_{\vec{p}}\cdot\vec{t}_{\vec{p}+\vec{k}}|\ll|\vec{t}_{\vec{p}}||\vec{t}_{\vec{p}+\vec{k}}|$. According to Ref.~\citep{CLee2025}, however, the odd-parity antiferromagnetic state with spin polarized Fermi surface can appear when $|\vec{t}_{\vec{p}}\cdot\vec{t}_{\vec{p}+\vec{k}}|^{2}-|\vec{t}_{\vec{p}}\times\vec{t}_{\vec{p}+\vec{k}}|^{2}$ is positive around the VHSs. Given $|\vec{t}_{\vec{p}}\cdot\vec{t}_{\vec{p}+\vec{k}}|^2+|\vec{t}_{\vec{p}}\times\vec{t}_{\vec{p}+\vec{k}}|^{2}=|\vec{t}_{\vec{p}}|^2|\vec{t}_{\vec{p}+\vec{k}}|^2$, it suggests a strong tension between the commensurability and the emergence of the spin-polarization over the Fermi surface: larger $f$- or $h$-wave spin-splitting in electronic structure is more compatible with an incommensurate phase than its commensurate counterpart when the nesting between type-II VHSs drives the antiferromagnetic instability.

\section{Effect of Spin-orbit coupling}\label{sec:SOC}

We have so far concentrated on the non-relativistic limit and discussed the symmetry-enforced and symmetry-assisted tendency towards incommensurate phases. We turn to discussing the effect of small SOC. Turning on the SOC, the spin- and real-space transformations
are locked together, which naturally lifts the degeneracy protected by the SO(3) spin rotation group. Here, we will focus on the effect of the SOC on systems that permit $f$- or $h$-wave AFM. For brevity, we consider systems where inversion is not a site symmetry; the case where inversion is a site symmetry is summarized in Appendix~\ref{app:SOC}.

In the non-symmorphic systems of interest to our work, at unit-cell doubling wavevector $\vec{Q}$, SOC will generically break down the six spin degrees of freedom into two two-dimensional mixed-parity irreps for in-plane moments and one for out-of-plane moments. Since the coplanar state can be generated by a single in-plane mixed-parity irrep, we only need to consider the splitting between the in- and out-of-plane magnetic moments. With this in mind, we make the simplifying assumption that the in-plane SO(2) spin-rotation symmetry is preserved at $\vec{Q}$ in the presence of SOC. In this case, the order parameters are expressed most naturally in terms of their in-plane $\vec{S}_{\tau,\parallel}=(S_{\tau,x}, S_{\tau,y} )$ and out-of-plane $S_{\tau,z}$ components. 
To first order in $\vec{q}$, we find that the GL free energy now includes the additional terms
\begin{align}
  \mathcal{F}_{\rm SOC} = 
    & \alpha_{\rm SOC} \sum_{\tau}(|\vec{S}_{\tau,\parallel}|^2 - |S_{\tau,z}|^2) \label{eq:GLwSOC}\\
    & + \sqrt{2} i\lambda_{\parallel z}\vec{q}\cdot\{ (\vec{S}_{A,\parallel} S_{A,z}^* - \vec{S}_{B,\parallel} S_{B,z}^*)-{\rm{c.c}}\},\nonumber
\end{align}
where $\vec{q}=(q_x, \pm q_y)$ or $(\pm q_y, q_x)$ depending on the representation chosen from Table~\ref{tab:D4h_tau_x}. The coefficient $\alpha_{\rm SOC}$ accounts for the splitting of the in- and out-of-plane magnetic orders at $\vec{Q}$. The sign of $\alpha_{\rm SOC}$ is of particular importance for the stability of the coplanar state: for $\alpha_{SOC} <0$ ($\alpha_{SOC} >0)$, SOC favours (disfavours) in-plane magnetic ordering relative to out-of-plane ordering, so the coplanar state is allowed (disallowed) as a possible instability. $\lambda_{\parallel z}$ is the coefficient of the pseudo-Lifshitz invariant, which is similar to the well-known Dzyaloshinskii–Moriya interaction \cite{Dzyaloshinskii1958,Moriya1960,Dzyaloshinskii1964DMI}. The SOC also introduces corrections which are quadratic in $\vec{q}$, e.g. a compass anisotropy~\cite{Farrell2014}, but we focus on leading-order effects here. In the small SOC limit, $\alpha_{\rm SOC}$ goes quadratically with the strength of SOC, whereas $\lambda_{\parallel z}$ is found to be linear in SOC. Despite the introduction of a linear in $\vec{q}$ term in the GL free energy, finite SOC does not guarantee incommensuration. This is because, in contrast to a true Lifshitz invariant, e.g. in Eq.~\eqref{eq:MomentumGL_w_Lifshitz}, the pseudo-Lifshitz term couples non-degenerate magnetic order parameters when $\alpha_{\rm SOC} \neq 0$. For small SOC, the instability condition derived in Sec.~\ref{sec:fwave} is modified to
\begin{align} \label{eq:SOCconstraint}
    D_{\rm Max} \equiv D -|K| - \frac{\lambda_{\parallel z}^2}{|\alpha_{\rm SOC}|}  < 0,
\end{align}
where $D_{\rm Max}$ is related to the curvature of the maximum eigenvalue of the susceptibility matrix. We see that the SOC tends to push the system towards incommensurate ordering when $\lambda^2_{\parallel z}$ is large compared to the splitting $|\alpha_{\rm SOC}|$.

We now consider an explicit tight-binding model which allows us to determine the coefficients in Eq.~\eqref{eq:GLwSOC} from a microscopic theory.  
Restricting our attention to locally noncentrosymmetric systems whose low-energy physics is governed by a single electronic orbital per site \cite{Fu2010,Youn2012,Yanase2016,Xie2020,Shishidou2021,Cavanagh2022}, we modify the tight-binding model in Eq.~\eqref{eq:TBHam} to
\begin{align}\label{eq:TBHamSOC}
    \hat{H}_{{\rm TB}} & =\sum_{\vec{k}}\hat{C}_{\vec{k}}^{\dagger}\{(\varepsilon_{\vec{k},0}\tau_{0}+\vec{t}_{\vec{k}}\cdot\vec{\tau}) \sigma_0 + \tau_z (\vec{\lambda}_{\vec{k}} \cdot \vec{\sigma}) \}\hat{C}_{\vec{k}},
\end{align}
where $\sigma_i$ denote the Pauli matrices in spin space and $\tau_i \sigma_j\equiv \tau_i \otimes \sigma_j$ is understood to be a Kronecker product. Note that inversion symmetry, $U_{\rm I} = \tau_x\sigma_0$, requires that $\epsilon_{\vec{k},0}$ and $t_{\vec{k},x}$ are even in $\vec{k}$, and $t_{\vec{k},y}$ and $\vec{\lambda}_{\vec{k}}$ are odd. When modeling tetragonal systems with Eq.~\eqref{eq:TBHamSOC}, the corresponding GL free energy for magnetic ordering exhibits an in-plane SO(2) spin rotation symmetry at the unit-cell doubling wavevector $\vec{Q}$, and thus takes the form given in Eq.~\eqref{eq:GLwSOC}. Explicit expressions for $\alpha_{\rm SOC}$ and $\lambda_{\parallel z}$ derived from this tight-binding model are provided in Appendix~\ref{app:SOC}.

In Fig.~\ref{fig:SOCsusceptibility}, we illustrate the effect of SOC using the tight-binding model Eq.~\eqref{eq:TBHamSOC}. In Fig.~\ref{fig:SOCsusceptibility}(a), we compare the in-plane, same-sublattice spin susceptibility $\chi^{xx}_{\tau\tau}(\vec{k})$ to the out-of-plane susceptibility $\chi^{zz}_{\tau\tau}(\vec{k})$ at unit-cell doubling vector $\vec{Q}=(\pi,\pi,0)$ as a function of $\mu$. Both susceptibilities are peaked for $\mu$ values which lead to strong interband nesting. We observe that the presence of SOC enhances $\chi^{xx}_{\tau\tau}(\vec{Q})$ relative to $\chi^{zz}_{\tau\tau}(\vec{Q})$ near the interband nesting condition, i.e. $\alpha_{\rm SOC}<0$. This is expected if $\frac{1}{2}\vec{\lambda}_{\vec{k},\parallel} \cdot \vec{\lambda}_{\vec{k}+\vec{Q},\parallel}<  \lambda_{\vec{k},z}\lambda_{\vec{k}+\vec{Q},z}$ holds where the interband nesting is significant, and follows from Eq.~\eqref{eq:deltaChi}. For systems described by Eq.~\eqref{eq:TBHamSOC}, assuming nearest-neighbor hopping dominates, the condition for SOC-enhanced in-plane ordering is likely satisfied. Therefore, if the coplanar state is the leading instability in the non-relativistic limit, we expect it to be robust to perturbations due to finite SOC. As we move from interband to intraband nesting, the splitting in the susceptibilities due to SOC changes sign, and so there generically exist values of $\mu$ such that $\chi^{xx}_{\tau\tau}(\vec{Q})=\chi^{zz}_{\tau\tau}(\vec{Q})$. Recalling the instability condition, Eq.~\eqref{eq:SOCconstraint}, we see that a transition between magnetic instabilities with in-plane and out-of-plane ordering must proceed through an incommensurate phase.

In Fig.~\ref{fig:SOCsusceptibility}(b), we calculate $D_{\rm Max}$ defined in Eq.~\eqref{eq:SOCconstraint} as a function of $\mu$. In the non-relativistic limit, $\lambda \rightarrow 0$, we see that $D_{\rm Max} = D - |K| > 0$ near the interband nesting condition $\mu \approx -0.3t$, and so the instability condition is not satisfied, i.e., we expect commensurate ordering. In contrast to the results presented in Fig.~\ref{fig:Driven_By_vHS}, this example illustrates a scenario where intersublattice hopping is insufficient to drive incommensuration. Including finite SOC, $\lambda=0.1t$, we see that $D_{\rm Max} <0$ for all $\mu$ considered, and thus, the system is unstable to incommensurate order. Following our previous discussion, we also note the divergence of $D_{\rm Max}$ where the splitting $\alpha_{\rm SOC}$ changes sign. 

Incommensuration driven by finite SOC is demonstrated in Figs.~\ref{fig:SOCsusceptibility}(c) and (d),  where we compute $\chi^{xx}_{\tau\tau}(\vec{k})$, $\chi^{zz}_{\tau\tau}(\vec{k})$, ${\rm Im} [\chi^{xz}_{\tau\tau}(\vec{k})]$ and the maximum eigenvalue of the susceptibility matrix, $\chi_{\rm Max}(\vec{k})$, along high-symmetry lines $\vec{k}=(u,u,0)$ and $\vec{k}=(u,\pi,0)$, respectively. We see that $\chi_{\rm Max}(\vec{k})$ in the non-relativistic limit, denoted by dashed lines, attains its maximum at $\vec{Q}=(\pi,\pi,0)$, leading to commensurate ordering.
In contrast, for finite SOC $\lambda=0.1t$, $\chi_{\rm Max}(\vec{k})$ is peaked away from $\vec{Q}=(\pi,\pi,0)$ due to contributions from spin-mixing terms, e.g. ${\rm Im} [\chi^{xz}_{\tau\tau}(\vec{k})]$, which disperse linearly near $\vec{Q}=(\pi,\pi,0)$. The maximum in the susceptibility is found to lie along the $\vec{k}=(u,\pi,0)$ line where the contribution from intersublattice hopping $\chi_{\rm AB}(\vec{k})$ is vanishing (see Sec.~\ref{sec:fwave}). This indicates that SOC is the primary mechanism driving incommensuration. Recalling the GL free energy in Eq.~\eqref{eq:GLwSOC}, this tendency towards incommensurate order is then attributed to the existence of a pseudo-Lifshitz invariant introduced by finite SOC. Additionally, since the pseudo-Lifshitz term couples the in- and out-of-plane magnetic moments, incommensurate order will be accompanied by a canting of the spins away from the in- or out-of-plane order favored by SOC at $\vec{k}=\vec{Q}$. 

\begin{figure}
\includegraphics[width=0.95\columnwidth]{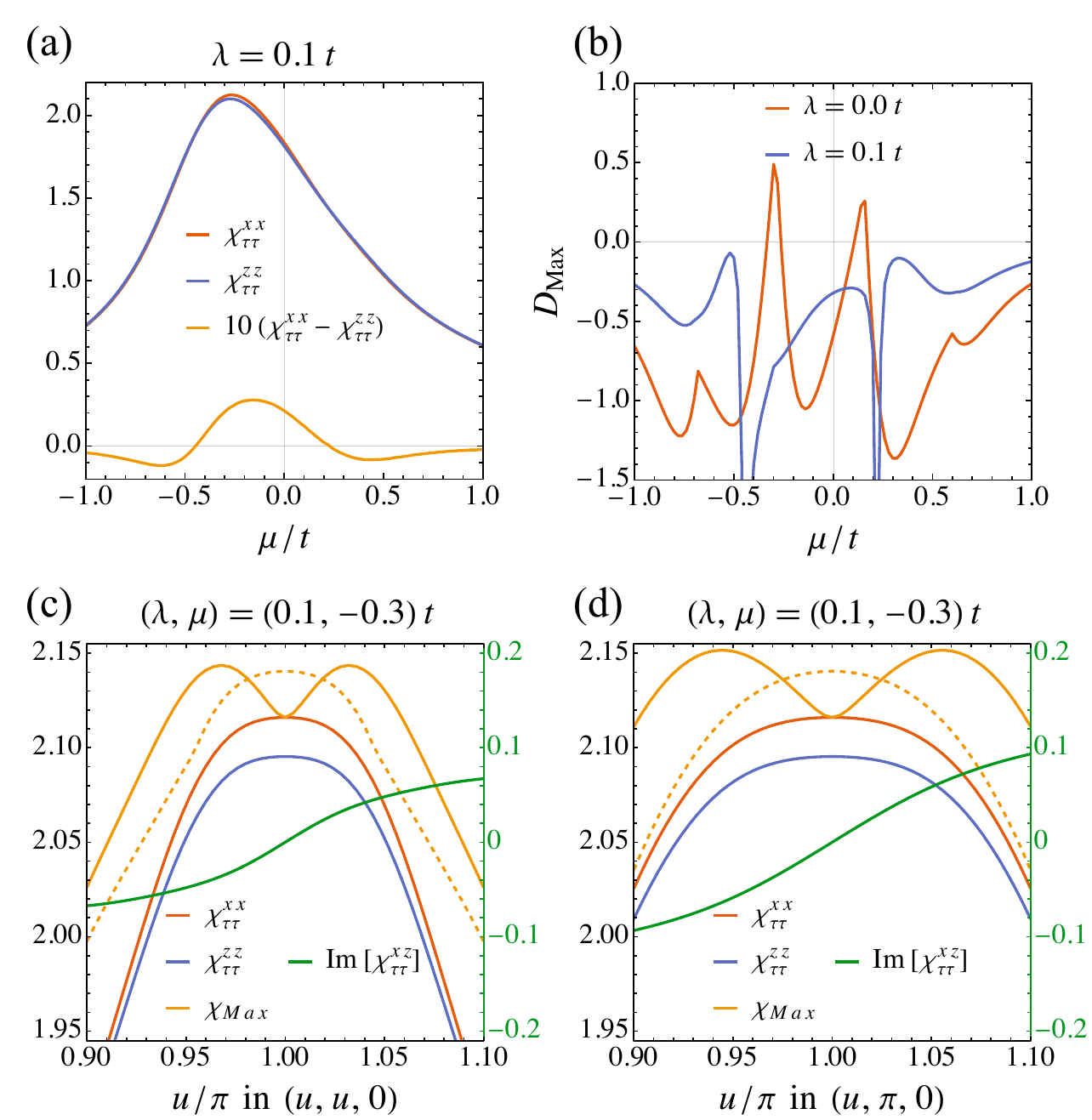}\caption{ (a) and (b) Comparison of the in-plane $\chi^{xx}_{\tau\tau}(\vec{k})$ and out-of-plane $\chi^{zz}_{\tau\tau}(\vec{k})$ susceptibilities, and calculation of $D_{\rm Max}$, respectively, as a function of chemical potential $\mu$ at unit-cell doubling wavevector $\vec{Q}=(\pi,\pi,0)$. (c) and (d) $\chi^{xx}_{\tau\tau}(\vec{k})$, $\chi^{zz}_{\tau\tau}(\vec{k})$, ${\rm Im}[\chi^{xz}_{\tau\tau}(\vec{k})]$ and maximum eigenvalue of the susceptibilty matrix, $\chi_{\rm Max}(\vec{k})$, along high symmetry lines $\vec{k}=(u,u,0)$ and $\vec{k}=(u,\pi,0)$, respectively, near $(\pi,\pi,0)$. The dashed lines denote $\chi_{\rm Max}(\vec{k})$ in the non-relativistic limit $\lambda =0.0t$ and ${\rm Im}[\chi^{xz}_{\tau\tau}(\vec{k})]$ is plotted against the
$y$-axis on right hand side. These calculations are carried out using the tight-binding model Eq.~\eqref{eq:TBHamSOC} with the same choice of $\varepsilon_{0,\vec{k}}$ and $\vec{t}_{\vec{k}}$ as Fig.~\ref{fig:Driven_By_vHS}, and $\vec{\lambda}_{\vec{k}}=(-\lambda \sin{k_y}, \lambda\sin{k_x},0)$. For these figures we set $(t_\perp,t^\prime)=(0.2,0.1)t$ and fix the temperature at $T=0.1t$. Note that we plot $D_{\rm Max}/10$ in (b) for $\lambda=0.1t$ to improve clarity. 
}\label{fig:SOCsusceptibility}
\end{figure}

\section{Discussion}

We have studied the stability of the transition into a unit-cell doubling AFM state with odd-parity spin-polarization pattern from the normal phase with respect to incommensuration.
In the case of a $p$-wave AFM, our symmetry analysis identifies that a Lifshitz invariant of non-relativistic origin exists if the space group of the normal state allows a continuous transition into a $p$-wave AFM state. The existence of a Lifshitz invariant destabilizes the continuous transition into the unit-cell doubling order from the normal state. We also demonstrate that the same conclusion can be drawn in the classical Heisenberg model in the non-relativistic limit in accordance with that the analysis is based on the symmetry of the order parameter. Therefore, the preference to an incommensurability to an unit-cell doubling order in the non-relativistic limit is generic.

Our result on the $p$-wave AFM state does not rule out its relevance as the ground state in non-symmorphic centrosymmetric systems, but this does place a significant constraint on the phase diagram of $p$-wave antiferromagnets: the transition into such a phase from the normal state should either pass through an incommensurate phase or be discontinuous.
The reported phase diagrams of CeNiAsO, RMnO$_{3}~$(R:Lu, Tm), and MnS$_2$~\citep{Shan2019CeNiAsO,Fangjun2023CeNiAsO,YuPomjakushin2009RMnO3,Mukherjee2017RMnO3,Chattopadhyay1991MnS2}, where unit-cell doubling antiferromagnetic states appear through a lock-in transition, are in agreement with our conclusion. We also note that the unit-cell doubling  antiferromagnetic state appearing in FeTe, which is subject to the non-symmorphic space group 129 (P4/nmm), emerges through a discontinuous transition with an ordering vector $\vec{Q}=(0,\pi,0)$ for which a Lifshitz invariant of non-relativistic origin prevents a continuous transition into this phase~\citep{Bao2009FeTe,Shiliang2009FeTe,Xu2009FeTe}.

In this respect, our result suggests that that the $p$-wave AFM with a unit-cell doubling $\vec{Q}$ is similar to the conventional heli- or spiral-magnets in terms of the propensity towards incommensuration in spite of its unit-cell doubling ordering vector. However, it can still provide advantageous venue for device application if spin-polarization over the Fermi surface persists, even if it is not perfectly odd-in-parity. Recent realization of electrical switching of the spin chirality in the helimagnet Nil$_2$ demonstrates this possibility~\citep{Song2025}.

For the odd-parity AFM which potentially exhibit $f$-wave or $h$-wave spin polarization patterns,
we consider the itinerant magnetism scenarios driven by the nesting
between VHSs or the nesting between electron and hole pockets. In
the first case, the symmetry condition for the existence of a mixed-parity
irrep at $\vec{Q}=(\pi,\pi,0)$ also enforces the VHSs to be positioned
off time-reversal invariant momentum, and thus to be turned into the
so-called type-II VHSs~\citep{Yao2015,CLee2025RG}. By using a Hubbard model with a tight-binding
model for generic two-band systems, we show that the incommensurate
nesting vector between the type-II VHSs makes the transition into an incommensurate state favored. 

Extending our analysis to locally noncentrosymmetric systems with weak SOC, we find that the SOC may favor AFM ordering with in-plane magnetic moments, allowing for coplanar ordering. However, SOC also allows for linear gradient terms, which couple magnetic order parameters in different irreps. These so-called pseudo-Lifshitz invariants tend to destabilize commensurate ordering. The possibility of SOC-driven incommensuration further constrains the emergence of a commensurate $f$- or $h$-wave AFM through a continuous transition. This result is of particular relevance to the unconventional superconductor CeRh$_2$As$_2$~\citep{Khim2021}. This compound features the Rashba SOC due to locally non-centrosymmetric structure and the type-II van Hove saddle points around the time-reversal invariant $X(0,\pi,0)$ point in the Brillouin zone~\citep{XChen2024PRX,BChen2024Expt,Wu2024Expt}. Recent renormalization group approach has shown that these altogether promote an incommensurate antiferromagnetic instability~\citep{CLee2025RG} in consistency with antiferromagnetic signatures observed in the muon spin resonance and nuclear magnetic resonance measurements~\citep{Ogata2023, Khim2024uSR}. In particular, just below the supposed magnetic transition temperature, the muon spin resonance exhibits a broad internal field power distribution, which is an indication of an incommensurate AFM phase~\citep{Blinc1981uSR,LPLe1991uSR,LPLe1993uSR,Blundell2004uSR}; at lower temperatures, however, clear oscillations in the signal appear, indicating a commensurate state. These experimental observations are consistent with the theory presented here.

Finally, we discuss the relation of $p$-wave AFM with the proposed $p$-wave Pomeranchuk instability of a Fermi liquid. The $p$-wave Pomeranchuk instability was postulated as a mechanism for odd-parity superconductivity~\citep{Born1948,Hellenes2024}. However, recent theoretical works showed that a second-order $p$-wave Pomeranchuk instability is unlikely due to the charge/spin conservation in Fermi liquid~\citep{Kiselev2017Pomeranchuck,Wu2018Pomeranchuck}. Meanwhile, $p$-wave AFM has been proposed to provide a way to circumvent the argument based on the conservation law because the inversion-breaking $p$-wave spin-polarization appears as a secondary order, or a by-product, of the emergent antiferromagnetism~\citep{Yue2025,Hellenes2024}. To this open question, our analysis points that the second-order $p$-wave AFM instability is still not probable, paradoxically prevented by the very non-symmorphic symmetries that allow it.

\acknowledgments
This work was supported by the Marsden Fund Council from Government funding, managed by Royal Society Te Ap\={a}rangi, Contract No. UOO2222.

\bibliography{ref}
\vfill\eject\clearpage

\appendix

\section{Real-valued two-dimensional mixed-parity irreducible representations in other space groups\protect{}}\label{App:Reality_MixedIrrep}
In the main text, we have focused on the real-valued mixed parity
irreducible representations (irreps) at unit-cell doubling vectors
$\vec{Q}$ whose little co-group is isomorphic to $D_{{\rm 2h}}$
or $D_{{\rm 4h}}$. The real-valued two-dimensional mixed parity irreps
and the symmetry-allowed spin-polarization in the odd-parity antiferromagnetic
phase are summarized in Tables~\ref{tab:LIandSPP_C2h_C4h}~to~\ref{tab:LIandSPP_Oh}, where all $p$-wave spin-polarization patterns imply the existence of a Lifshitz invariant.

\begin{table}[h]
\centering
\begin{tabular}{ccccc}
\toprule 
\multirow{3}{*}{$C_{{\rm 2h}}$} & ${\cal I}$ & $C_{2z}$ & LI & SPP\tabularnewline
\cmidrule{2-5}
 & \multirow{2}{*}{$\tau_{x}$} & $i\tau_{y}$ & $\partial_{z}$ & $p_{z}$\tabularnewline
 &  & $\tau_{z}$ & $\partial_{x},\partial_{y}$ & $p_{x},p_{y}$\tabularnewline
\midrule
\multirow{3}{*}{$C_{{\rm 4h}}$} & ${\cal I}$ & $C_{4z}$ & LI & SPP\tabularnewline
\cmidrule{2-5}
 & \multirow{2}{*}{$\tau_{x}$} & $i\tau_{y}$ & $\partial_{z}$ & $p_{z}$\tabularnewline
 &  & $\tau_{z}$ & No & $p_{x}p_{y}p_{z},(p_{x}^{2}-p_{y}^{2})p_{z}$\tabularnewline
\bottomrule
\end{tabular}\caption{Lifshitz invariant allowed by real-valued two-dimensional mixed-parity
irreps at a unit-cell doubling $\vec{Q}$ whose little co-group is
isomorphic to $C_{{\rm 2h}}$ or $C_{{\rm 4h}}$ along with the spin-polarization pattern in
the associated non-collinear coplanar antiferromagnetic phase. The
representation matrix for the inversion is chosen as $\tau_{x}$.
LI: Lifshitz invariant, SPP: spin-polarization pattern.}\label{tab:LIandSPP_C2h_C4h}
\end{table}

\begin{table}
\begin{centering}
\begin{tabular}{cccccc}
\toprule 
\multirow{2}{*}{$D_{{\rm 3d}}$} & ${\cal I}$ & $C_{3z}$ & $C_{2x}$ & LI & SPP\tabularnewline
\cmidrule{2-6}
 & \multirow{1}{*}{$\tau_{z}$} & $\tau_{0}$ & $\tau_{x}$ & $\partial_{z}$ & $p_{z}$\tabularnewline
\midrule
$C_{{\rm 6h}}$, $D_{{\rm 6h}}$ & ${\cal I}$ & $C_{6z}$ & $C_{2x}$ & LI & SPP\tabularnewline
\midrule
$\vec{\tau}_{6z}=0$ & \multirow{2}{*}{$\tau_{z}$} & $\tau_{0}$ & $\tau_{x}$ & $\partial_{z}$ & $p_{z}$\tabularnewline
\cmidrule{1-1}
$\vec{\tau}_{6z}=\hat{z}/2$ &  & $i\tau_{y}$ & $\tau_{z}$/$\tau_{x}$ & $\partial_{z}$ & $p_{z}$\tabularnewline
\bottomrule
\end{tabular}
\par\end{centering}
\centering{}\caption{Possible real-valued two-dimensional mixed-parity irreducible representations
at $\vec{Q}$ subject to $D_{3d}$, $C_{{\rm 6h}}$, or $D_{{\rm 6h}}$.
$\tau_{6z}$ denotes the half-translation associated with the six-fold
rotation $C_{6z}$. $\tau_{z}/\tau_{x}$ for $C_{2x}$ in the lower
block for $C_{{\rm 6h}}$ and $D_{{\rm 6h}}$ means that the representation
of $C_{2x}$ depends on whether its half-lattice translation $\vec{\tau}_{\perp}$
is $0$ or $\hat{z}/2$, respectively.
The last row also shows the mixed-parity irreps at $\vec{Q}$ subject to $C_{{\rm 6h}}$, where the lack of $C_{2x}$ in this group is to be understood.}\label{tab:LIandSPP_D3d_to_D6h}
\end{table}
\begin{table}
\begin{centering}
\begin{tabular}{cccccc}
\toprule 
${\cal I}$ & $C_{2y}$ & $C_{2,110}$ & $C_{3,111}$ & LI & SPP\tabularnewline
\midrule
\multirow{1}{*}{$\tau_{x}$} & $\tau_{0}$ & $\tau_{z}$ & $\tau_{0}$ & No & $p_{x}p_{y}p_{z}$\tabularnewline
\bottomrule
\end{tabular}
\par\end{centering}
\centering{}\caption{Possible real-valued two-dimensional mixed-parity irreducible representations
at $\vec{Q}$ subject to $O_{{\rm h}}$, which occurs at the $R(\pi,\pi,\pi)$
point in space groups No. 222 and 223, and the $H(2\pi,2\pi,2\pi)$
point in the space group No. 230.}\label{tab:LIandSPP_Oh}
\end{table}

\section{Further discussion of SOC} \label{app:SOC}

In this appendix, we first discuss the form of $\mathcal{F}_{\rm SOC}$ in Eq.~\eqref{eq:GLwSOC} for systems where inversion in a site symmetry. We then give explicit expressions for the spin splitting $\alpha_{\rm SOC}$ and the coefficient of the pseudo-Lifshitz invariant $\lambda_{\parallel z}$ in Eq.~\eqref{eq:GLwSOC} for the tight-binding model in Eq.~\eqref{eq:TBHamSOC}. 

\subsection{$\mathcal{F}_{\rm SOC}$ when inversion is a site symmetry}
The form of Eq.~\eqref{eq:GLwSOC} is constrained by the representation matrices presented in Tab.~\ref{tab:D4h_tau_x}. If we wish $\vec{S}_\tau$ to represent the magnetic moments localized on the sublattice labelled by $\tau$, the choice of $\tau_x$ as the representation of the inversion operator then corresponds to the case where inversion is not a site symmetry. This case is discussed in Sec.~\ref{sec:SOC} of the main text. We can determine the form of $\mathcal{F}_{\rm SOC}$ for the case where inversion is a site symmetry by applying a unitary transformation $U$ such that $\tau_x \rightarrow U \tau_x U^\dagger=\tau_z$, for example, $U=\exp(i\pi \tau_y /2 ) $. Applying the corresponding transformation to the magnetic order parameters $\vec{S}_\tau \rightarrow \sum_{\tau'} U_{\tau\tau'} \vec{S}_{\tau'}$, we find that $\mathcal{F}_{\rm SOC}$ transforms as
\begin{align} \label{eq:SOCclass2}
    \mathcal{F}_{\rm SOC} \rightarrow   \Tilde{\mathcal{F}}_{\rm SOC} =&\alpha_{\rm SOC} \sum_{\tau}(|\vec{S}_{\tau,\parallel}|^2 - |S_{\tau,z}|^2) \\
    & + \sqrt{2} i\lambda_{\parallel z}\vec{q}\cdot\{ (\vec{S}_{A,\parallel} S_{B,z}^* + \vec{S}_{B,\parallel} S_{A,z}^*)-{\rm{c.c}}\}, \nonumber
\end{align}
where the pseudo-Lifshitz term now introduces {\it intersublattice} coupling between in- and out-of-plane magnetic moments. We note that, while Eq.~\eqref{eq:GLwSOC} and Eq.~\eqref{eq:SOCclass2} differ in their details, the qualitative effects of the SOC discussed in Sec.~\ref{sec:SOC} remain valid. In particular, the instability condition Eq.~\eqref{eq:SOCconstraint} is unchanged.

\subsection{Explicit expressions for $\alpha_{\rm SOC}$ and $\lambda_{\parallel z}$}
We now derive the coefficients that appear in Eq.~\ref{eq:GLwSOC} in the case where the microscopic degrees of freedom are described by the tight-binding model given in Eq.~\eqref{eq:TBHamSOC}. First, we rewrite the tight-binding model as
\begin{align}
    \hat{H}_{{\rm TB}} & =\sum_{\vec{k}}\hat{C}_{\vec{k}}^{\dagger} \mathcal{H}_{\vec{k}} \hat{C}_{\vec{k}}, \\ \mathcal{H}_{\vec{k}}&=\varepsilon_{0,\vec{k}} \tau_0\sigma_0 + \vec{\varepsilon}_{\vec{k}} \cdot \vec{\gamma},
\end{align}
where $\vec{\varepsilon}_{\vec{k}} = (\vec{t}_{x,\vec{k}}, t_{y,\vec{k}}, \lambda_{x,\vec{k}}, \lambda_{y,\vec{k}}, \lambda_{z,\vec{k}})$ and $\vec{\gamma} = (\tau_x\sigma_0,\tau_y\sigma_0,\tau_z\sigma_x,\tau_z\sigma_y,\tau_z\sigma_z)$. Since $\vec{\gamma}$ is a vector of mutually anticommuting matrices, the eigenvalues of $\mathcal{H}_{\vec{k}}$ are simply given by $\xi_{\vec{k},\pm} = \varepsilon_{0,\vec{k}} \pm |\vec{\varepsilon}_{\vec{k}}|$.

Following the discussion in Sec.~\ref{sub:VHS}, using a field theoretic approach, we may derive the Landau free energy in terms of magnetic order parameters $\vec{S}_\tau(\vec{k})$. We first consider the splitting $\alpha_{\rm SOC}$ due to SOC for commensurate ordering.  To second order in $\vec{S}_\tau$, the qualitatively new terms introduced to the free energy by SOC 
\begin{align}
    \mathcal{F}_{\rm SOC} = -\sum_{\vec{k}}\sum_{\tau\tau'} \vec{S}_\tau(-\vec{k}) \cdot \delta\chi_{\tau\tau'} (\vec{k}) \cdot  \vec{S}_{\tau'}(\vec{k}).
\end{align}
where $\delta\chi(\vec{k})$ is the corresponding correction to the susceptibility matrix. Note that the inclusion of SOC will also generically introduce numerical corrections to the $D$ and $K$ coefficients appearing in the GL free energy Eq.~\eqref{eq:fwaveGLfree}. The splitting $\alpha_{\rm SOC}$ is determined by $\delta \chi(\vec{k}=\vec{Q})$ which is given by 
\begin{align}\label{eq:deltaChi}
    \delta\chi_{\tau\tau'}^{ij}(\vec{Q})= \delta_{\tau\tau'}\sum_{\vec{p}}\sum_{a,a'=\pm} a a' \chi_{aa'}  \hat{\lambda}_{\vec{p},i} \hat{\lambda}_{\vec{p}+\vec{Q},j},
\end{align}
where $\chi_{aa'}$ is the multiband Lindhard function given in Eq.~\eqref{eq:lindhard} evaluated at $\vec{k}=\vec{Q}$ and  $\hat{\lambda}_{\vec{p}}=\vec{\lambda}_{\vec{p}}/|\vec{\varepsilon}_{\vec{p}}|$. $\delta\chi(\vec{Q})$ is diagonal in sublattice- and spin-space, and the splitting between in- and out-of-plane ordering is given by 
\begin{align}
   \alpha_{\rm SOC}=-\frac{1}{2}\left(\delta\chi^{xx}_{\tau\tau}(\vec{Q}) -\delta\chi^{zz}_{\tau\tau}(\vec{Q}) \right) .
\end{align}
For the case of Rashba SOC $\vec{\lambda}_{\vec{p}}=(-\lambda \sin{p_y}, \lambda\sin{p_x},0)$ and $\vec{Q}=(\pi,\pi,0)$, as considered in Fig.~\ref{fig:SOCsusceptibility}, we find that $\alpha_{\rm SOC} <0$ $(\alpha_{\rm SOC} >0)$ for interband (intraband) nesting.

We can determine $\lambda_{\parallel z}$ by expanding the susceptibility around the $\vec{Q}$ point. We have 
\begin{widetext}
    \begin{align}
    \frac{\partial \delta\chi(\vec{Q}+\vec{q})}{\partial q_i} \Big |_{\vec{q}\rightarrow 0} = i \tau_z \sum_{\vec{p}}\sum_{a,a'=\pm} a\frac{\frac{\xi_{\vec{p}+\vec{Q},a'}-\xi_{\vec{p},a}}{2T}{\rm sech}^{2}\frac{\xi_{\vec{p}+\vec{Q},a'}}{2T}-\tanh\frac{\xi_{\vec{p}+\vec{Q},a'}}{2T}+\tanh\frac{\xi_{\vec{p},a}}{2T} }{(\xi_{\vec{p},a}-\xi_{\vec{p}+\vec{Q},a'})^{2}}(\partial_{i}\xi_{\vec{p}+\vec{Q},a'})A_{\vec{p}}
\end{align}
\end{widetext}
where $A_{\vec{p}}^{ij} = \frac{1}{2} \sum_{k=x,y,x} \epsilon_{ijk} \hat{\lambda}_{\vec{p},k}$ and $\epsilon_{ijk}$ denotes the totally antisymmetric tensor. Here, we have assumed that $\vec{\lambda}_{\vec{p}}$ and $\vec{\lambda}_{\vec{p}+\vec{Q}}$ are odd in $\vec{p}$. The coefficient for the pseudo-Lifshitz term is then given by 
\begin{align}
    \lambda_{\parallel z} = \frac{i}{\sqrt{2}} \frac{\partial \delta\chi_{AA}^{xz}(\vec{Q}+\vec{q})}{\partial q_x} \Big |_{\vec{q}\rightarrow 0} 
\end{align}

\clearpage
\end{document}